# Analysis of Cohesive Micro-Sized Particle Packing Structure Using History-Dependent Contact Models

Raihan Tayeb, Xin Dou, Yijin Mao, Yuwen Zhang[1]

*Department of Mechanical and Aerospace Engineering*

*University of Missouri*

*Columbia, Missouri, 65211, USA*

**Abstract**

Granular packing structures of cohesive micro-sized particles with different sizes and size distributions, including mono-sized, uniform and Gaussian distribution, are investigated by using two different history dependent contact models with Discrete Element Method (DEM). The simulation is carried out in the framework of LIGGGHTS which is a DEM simulation package extended based on branch of granular package of widely used open-source code LAMMPS. Contact force caused by translation and rotation, frictional and damping forces due to collision with other particles or container boundaries, cohesive force, van der Waals force, and gravity are considered. The radial distribution functions (RDFs), force distributions, porosities, and coordination numbers under cohesive and non-cohesive conditions are reported. The results indicate that particle size and size distributions have great influences on the packing density for particle packing under cohesive effect: particles with Gaussian distribution have the lowest packing density, followed by the particles with uniform distribution; the particles with mono-sized distribution have the highest packing density. It is also found that cohesive effect to the system does not significantly affect the coordination number that mainly depends on the particle size and size distribution. Although the magnitude of net force distribution is different, the results for porosity, coordination number and mean value of magnitude of net force do not vary significantly between the two contact models.

**Keywords:** Granular Packing; Discrete Element Method; Size distribution; Radial Distribution Function

**Nomenclature**

*d*     diameter of particle, m

*e*     coefficient of restitution

*g*     gravity, m/s$^2$

*m*    mass of particle, kg

*v*     velocity, m/s

---

[1] Corresponding author. Email: zhangyu@missouri.edu.

| | |
|---|---|
| $F$ | force on particle, N |
| $I$ | moment of inertia, kg·m$^2$ |
| $R$ | radius of particle, m |
| $T$ | torque, N·m |
| $X$ | position vector, m |
| $Y$ | Young's modulus, Pa |

**Greek Symbols**

| | |
|---|---|
| $\gamma$ | damping coefficient, s |
| $\theta$ | rotational angle, rad |
| $\mu_s$ | sliding friction coefficient |
| $\mu_r$ | rolling friction coefficient |
| $\xi_n$ | normal direction displacement, m |
| $\xi_t$ | tangential displacement, m |
| $\rho$ | particle density, kg/m$^3$ |
| $\sigma$ | standard deviation |
| $\sigma_P$ | Poisson ratio |
| $\omega$ | angular velocity, rad/s |

**Introduction**

    Granular packing simulation is usually used to model structures of materials that are involved in many industrial applications ranging from manufacturing raw materials to developing advanced products. The impact of particle properties on their packing structures is of the prime importance to the entire packing process and is always essential for fabrication. A better understanding of packing is beneficial to optimize and to improve the industrial applications. This topic has been intensively studied in the past decades; many of them focused on the micro level packing [1-3] where packing density, which is equals to unity minus its porosity, is used as their main indicator to measure and evaluate quality of packing structure [4]. Among those works, researchers, by varying particle sizes, size distributions or forces involved in the packing process, obtained detailed information of packing structures and revealed weighted influences from different parameters [5-9]. Some of them are interested in cohesive effect that is substantially caused by cohesive forces such as van der Waals force, capillary force that is associated with wet particles and electrostatic force that can be important for finer particles. Cohesive effect turns out to be of importance in particular situations, for example, when packing containers are no longer rigid but are kind of material that has similar properties like dry sand, cement or wet soil, or it is not even

solid just like settling particles in the fluid where effect of gravity will reduce and cohesive effect will become significant [10-12].

The studies on behaviors of cohesive particles are usually carried out by changing particle sizes, mixture component percentage if particles are not made with the same material or fluid density if particles are settled down into a fluid. Effects of particle diameters (mean diameters for mixed particle cases) are considered to be a great factor that influences the packing structures thereby worth more attention. Boundary condition is another important factor that alters the force and deformation of the packing structure. Previously, researchers mostly adopted the periodic boundary conditions for the packing process where particles that exit the simulation box will come back in opposite direction in order to maintain the number of particles in the simulation box [13]. This approach allows that the simulation can be carried out smoothly, because of lower chances of losing systemic energy and generating huge interactive force.

In this work, the cohesive effects associated with size distributions, which include mono-sized, uniform, and Gaussian distributions, will be investigated by using two different history-dependent contact models. It is worth to point out that the uniform distribution of particle size indicates the sizes of particles linearly increase from the minimum to the maximum. The range for uniform distribution is kept at a constant of 40 μm. For Gaussian distribution, the STD (standard deviation), $\sigma$, is set as 13.33 μm for all cases. In addition, fixed boundary conditions are applied to all sides of the simulation box for all the cases such that the particles may collide with boundaries during the packing process which will cause energy losses due to friction. In order to understand the fundamentals that govern the cohesive particle packing, a series of well-designed programs are developed based on the Discrete Element Method [14-17]. LIGGGHTS[18] that is based on LAMMPS [19], providing a simulator of solving particle related problems from industrial applications, is employed to resolve the packing process. The simulation results, including radial distribution function (RDF) that indicates how number density changes with distance from a selected reference point, force distributions that give a view on the magnitude of forces acting on the particles, porosities and coordination numbers, are presented in this paper.

**Numerical Methods and Physical Models**

It is well known that any motions of a rigid particle can be decomposed to two parts: translational and rotational motions. Referring to the Newton's second law, the governing equations for each particle during this packing process can be written as:

$$m_i \frac{\partial^2 \mathbf{X}_i}{\partial t^2} = \mathbf{F}_i \tag{1}$$

$$I_i \frac{d^2 \boldsymbol{\theta}_i}{dt^2} = \mathbf{T}_i \tag{2}$$

$$\mathbf{F}_i = \mathbf{F}^n_{ij} + \mathbf{F}^t_{ij} \tag{3}$$

$$\mathbf{T}_i = \mathbf{T}^r_{ij} + \mathbf{T}^t_{ij} \tag{4}$$

where $m_i$ is the mass of the i$^{th}$ particle, $\mathbf{X}_i$ is the position vector of the i$^{th}$ particle, $I_i$ is the moment of inertia that equals to $0.4 m_i R_i^2$ and the rotated angle of particle i is represented by $\boldsymbol{\theta}_i$. The symbol $\mathbf{F}_i$ in Eq. (1) is the resultant contact force generated by two collided particles, i and j. This force can be decomposed further into two components: one is contact force in normal direction $\mathbf{F}^n_{ij}$ and the other is contact force in tangential direction $\mathbf{F}^t_{ij}$, as shown in Eq. (3). The symbol $\mathbf{T}_i$ in Eq. (2) is the resultant torque acting on the i$^{th}$ particle. It can also be decomposed into two components: torques caused by rolling friction and tangential force, respectively, as given in Eq. (4).

The two contact models adopted in this work are both history deformation dependent. The difference is from the relationship between deformation and contact force. Gran-Hertz-History model describe a nonlinear relationship between contact force and overlap distance, while Gran-Hooke-History model gives a linear relationship. The open-source software package LIGGGHTS provides both of these models. However the Gran-Hertz-History model is modified to include van der Waals force, which can be significant for small particles, and thereby refer to as Modified Gran-Hertz-History model.

The normal contact force $\mathbf{F}^n_{ij}$ can be determined by [20-22],

$$\mathbf{F}^n_{ij} = \left[ K_n \xi_n - \gamma_n \left( \mathbf{v}_{ij} \cdot \mathbf{n}_{ij} \right) \right] \mathbf{n}_{ij} \tag{5}$$

where in Modified Gran-Hertz-History model, parameters are given by

$$K_n = \frac{4}{3} Y_{eff} \sqrt{\bar{R} \xi_n}, \gamma_n = 2\sqrt{\frac{5}{6}} \beta_{eff} \sqrt{S_n m_{eff}}, \beta_{eff} = \frac{\ln(e)}{\sqrt{\ln^2(e) + \pi^2}}, S_n = 2 Y_{eff} \sqrt{\bar{R} \xi_n}.$$

and in Gran-Hooke-History model,

$$K_n = \left( \frac{16}{15} \sqrt{\bar{R}} Y_{eff} \right)^{0.8} \left( m_{eff} v_{ch}^2 \right)^{0.2}, \gamma_n = \sqrt{\frac{4 m_{eff} K_n}{1 + \frac{\pi^2}{\ln^2(e)}}}.$$

and $\mathbf{v}_{ij}$ represents the velocity of the particle i relative to velocity of the particle j, $\mathbf{n}_{ij}$ is the unit vector point from particle i to particle j, $e$ is the coefficient of restitution of the particles, $\bar{R} = R_i R_j / (R_i + R_j)$ is the effective radius that represent the geometric mean diameter of the i and j particle, $Y_{eff} = 1 / \left( \frac{1-\sigma_1^2}{Y_1} + \frac{1-\sigma_2^2}{Y_2} \right)$ is the effective Young's modulus that is calculated in terms of individual Young's modulus and Poisson ratio accordingly, $\xi_n = R_i + R_j - |R_{ij}|$ is the overlap in normal direction and $m_{eff} = \frac{m_i m_j}{m_i + m_j}$ is the effective masses of the particles. Characteristic velocity $v_{ch}$ is taken as unity in Gran-Hooke-History model.

The contact force in tangential direction is calculated by [23],

$$\mathbf{F}^t_{ij} = -\min\left[ \mu \left| \mathbf{F}^n_{ij} \right|, K_t \left( \boldsymbol{\xi}_t \cdot \mathbf{t}_{ij} \right) - \gamma_t \left( \mathbf{v}_t \cdot \mathbf{t}_{ij} \right) \right] \mathbf{t}_{ij} \tag{6}$$

where in Modified Gran-Hertz-History model, parameters are determined by,

$$K_t = 8 G_{eff} \sqrt{\bar{R} \xi_n}, \gamma_t = 2\sqrt{\frac{5}{6}} \beta_{eff} \sqrt{S_t m_{eff}}, S_t = 8 G_{eff} \sqrt{\bar{R} \xi_n},$$

$$G_{eff} = 1 / \left[ \frac{2(2-\sigma_1)(1+\sigma_1)}{Y_1} + \frac{2(2-\sigma_2)(1+\sigma_2)}{Y_2} \right].$$

and in Gran-Hooke-History model, $K_t = K_n$, $\gamma_t = \gamma_n$

$\xi_t = \int_{t_0}^{t} v_t dt$ represents the tangential displacement vector between the two spherical particles, $\mathbf{v}_t = [(\mathbf{v}_i - \mathbf{v}_j) \cdot \mathbf{t}_{ij}]\mathbf{t}_{ij} + (\boldsymbol{\omega}_i \times \mathbf{R}_i - \boldsymbol{\omega}_j \times \mathbf{R}_j)$ is the tangential relatively velocity, $t_{ij}$ is the unit vector along the tangential direction, $t_0$ is the time when the two particles just touch and have no deformation, t is the time of collision, $\omega_i$ or $\omega_j$ is the angular velocities of particles i or j and $R_i$ or $R_j$ is the vector running from the center of particle i or j to the contact point of the two particles.

The cohesive force is included in both Modified Gran-Hertz-History model and Gran-Hooke-History model. For the cohesive force, Johnson-Kendall-Roberts (JKR) model [24] based on Hertz elastic theory is used to estimate the cohesive behavior of the particles. In Hertz elastic theory, the normal pushback force between two particles is proportional to the area of overlap between the particles. Based on Hertz elastic assumption and meanwhile considering the contact surface as perfectly smooth, the JKR model here is satisfactorily accurate to determine the cohesive force. In fact, the basic idea is that if two particles are in contact, it adds an additional normal force tending to maintain the contact,

$$|\mathbf{F}| = kA \tag{7}$$

where $k$ is the surface energy density and $A$ is the particle contact area. For sphere-sphere contact [25], contact area $A$ is evaluated by,

$$A = \frac{\pi}{4} \times \frac{(dist - R_i - R_j)(dist + R_i - R_j)(dist - R_i + R_j)(dist + R_i + R_j)}{dist^2} \tag{8}$$

where *dist* is the central distance between the i and j particles. $R_i$ and $R_j$ are the radius of the $i^{th}$ and $j^{th}$ particle, respectively.

The van der Waals forces among particles are included only in the Modified Gran-Hertz-History model. The van der Waals force, $\mathbf{F}^v_{ij}$ between particles i and j is given by [26],

$$\mathbf{F}^v_{ij} = -\frac{H_a}{6} \times \frac{64 R_i^3 R_j^3 (h + R_i + R_j)}{(h^2 + 2R_i h + 2R_j h)^2 (h^2 + 2R_i h + 2R_j h + 4R_i R_j)^2} \tag{9}$$

where $H_a$ is the Hamaker constant, and $h$ is the separation of surfaces along the line of the centers of particles i and j. A minimum separation distance $h_{min}$ is considered to prevent $\mathbf{F}^v_{ij}$ becoming infinity when h goes to zero. The Hamaker constant is related to the surface energy density by [27]:

$$H_a = 24\pi k h_{min}^2 \tag{10}$$

The torque due to tangential contact force and the torque due to rolling friction are calculated in the same way for both models [28]:

$$\mathbf{T}^t_{ij} = \mathbf{R}_i \times \mathbf{F}^t_{ij} \tag{11}$$

$$\mathbf{T}^r_{ij} = \mu_r \bar{R} K_n \xi_n \frac{\boldsymbol{\omega}_{ij} \cdot \mathbf{t}_{ij}}{|\boldsymbol{\omega}_{ij}|} \mathbf{t}_{ij} \tag{12}$$

where $\boldsymbol{\omega}_{ij} = \boldsymbol{\omega}_i - \boldsymbol{\omega}_j$ is the relative angular velocity.

Table 1 shows the material properties and other physical coefficients used in these packing simulations. The material properties of the particles are same as those for iron. The surface energy density is calculated from the Hamaker constant. The material properties for the container is same as that for the particles.

Table 1 Values of the parameters used in the simulation process

| Parameters | Values |
|---|---|
| Particle density ρ | 7870 kg/m$^3$ |
| Young's modulus Y | $200\times10^9$ N/m$^2$ |
| Restitution coefficient e | 0.75 |
| Sliding friction coefficient $\mu_s$ | 0.42 |
| Rolling friction coefficient $\mu_r$ | $2\times10^{-4}$ |
| Poison ratio $\sigma_P$ | 0.29 |
| Hamaker constant, $H_a$ | $21.1\times10^{-20}$ J |
| Minimum separation distance, $h_{min}$ | $1\times10^{-10}$ m |
| Surface energy density, k | 0.280 J/m$^2$ |

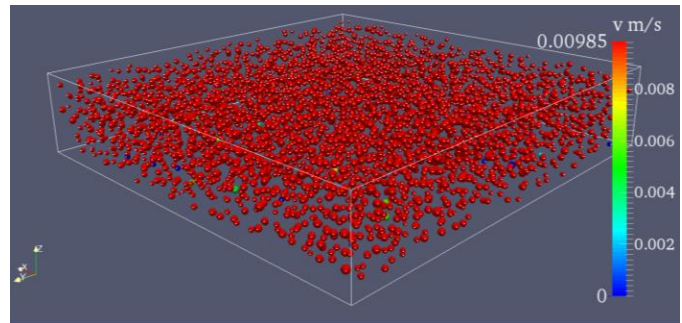

(a) Particles at t = $1\times10^{-8}$ sec

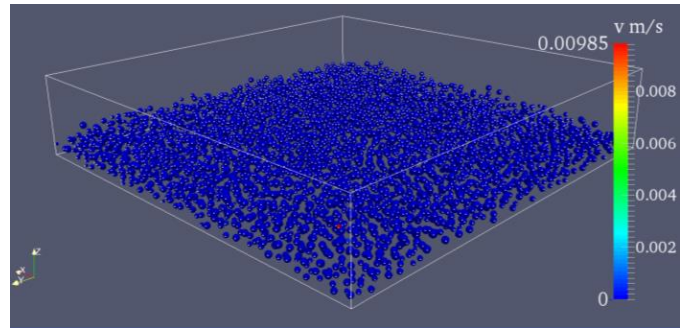

(b) Particles at t = 0.2 sec

Figure 1　Initial and final structure for Gaussian particles from Modified Gran-Hertz-History model with cohesion.

For each simulation, 4,500 particles are settled in a simulation box having length and width equal to 0.006m and the particles have no initial physical contact among them. The initial porosity is kept constant at 0.75. Figure 1 shows the initial state of Gaussian particle packing. As the simulation time increases, the particles begin to fall down due to gravity and then collide with other particles or with the boundaries. In this work, all six sides of the simulation box are considered as physically stationary. Considering the fact that the contact force is mainly related to the particle deformation, the time-step must be sufficiently small to prevent any unrealistic overlap [29]. In this work, the time step is set to be $1\times10^{-8}$ s for all simulation cases. It should be pointed out that the velocity of each particle will hardly reach zero completely but the magnitude of velocity will approach to an extremely small value. In this work, the particles are considered to be completely stationary when their mean velocities are below $1\times10^{-8}$ m/s. The results are presented through three parameters which are widely used to measure the packing structure: (1) radial distribution function (RDF), (2) porosity that is the ratio of total volume of void space to the volume taken by all the particles, and (3) coordination number that is defined as the number of particles that are contacting with the one chosen as reference center.

**Results and Discussions**

Sixty scenarios are studied in this work: five different mean radius (75μm, 85 μm, 100 μm, 110 μm, and 120 μm) and three different size distributions (mono-sized, uniform and Gaussian) for two contact models (Modified Gran-Hertz-History model and Gran-Hooke-History model) with and without cohesion. The results are presented in the form of porosity, coordination numbers, RDF and force distribution, when the all particles are completely packed. Forces considered in the packing process include contact force, which is decomposed into normal and tangential components, viscoelastic and frictional forces generated when collision occurs, gravity which drives the particles to fall down, cohesive force and van der Waals force which are considered as external forces acting on themselves. Figures 1-4 show the initial and final packing structures for these three distributions with mean diameter of 75 μm. It should be noted that the deformation calculation is very important for packing simulation since the oversimplified model of calculating overlap distance is always the main reason that leads to the simulation crash by introducing unrealistic energy. Two basic rules are applied to these packing simulations: one is that particles are always considered as rigid body even though a deformation is considered by the chosen model, and the other is that the critical central distance is set for particle deformation. The critical distance is $1.01(d_1+d_2)/2$ where $d_1$ and $d_2$ are the diameters of the two particles. It means when the central distance of two particles is less than the critical distance the two particles are considered to be in direct contact [6].

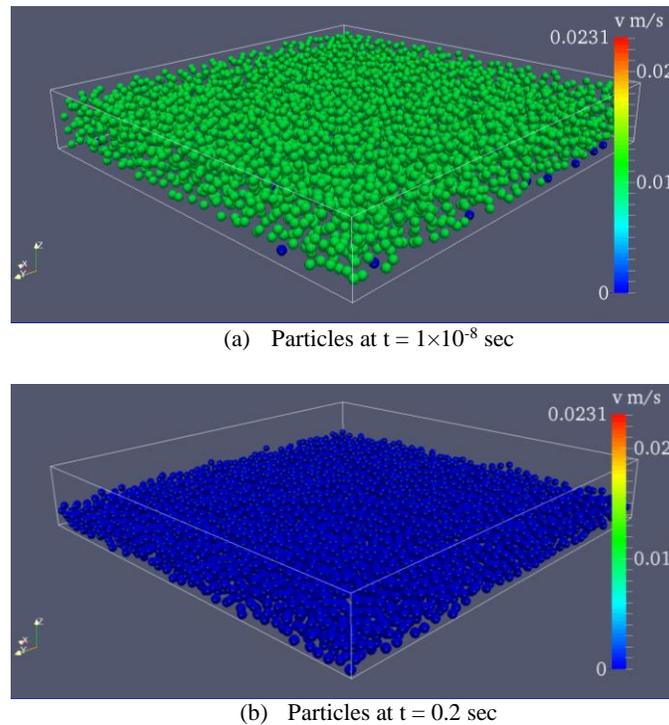

(a) Particles at t = 1×10$^{-8}$ sec

(b) Particles at t = 0.2 sec

Figure 2    Initial and final packing structure for mono-sized particles from Modified Gran-Hertz-History model with cohesion.

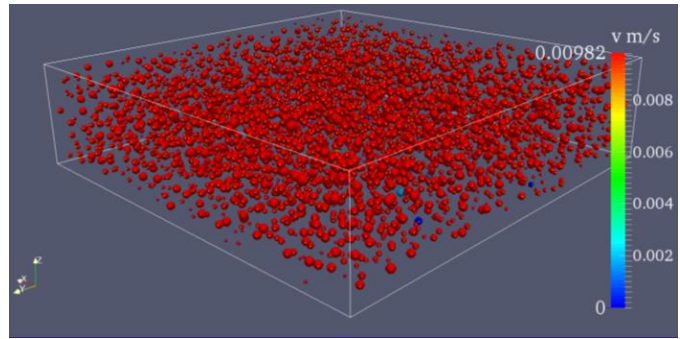
(a) Particles at t = 1×10⁻⁸ sec

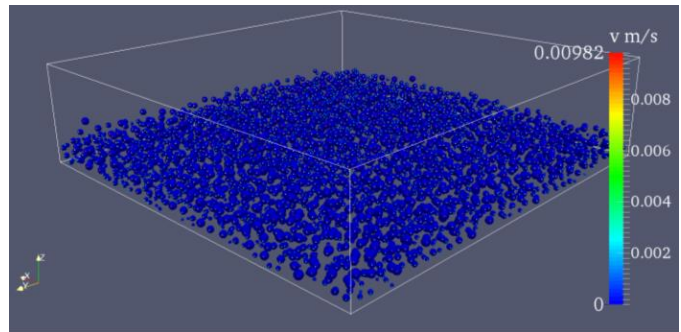
(b) Particles at t = 0.2 sec

Figure 3     Initial and final packing structure for uniform size particles from Modified Gran-Hertz-History with cohesion.

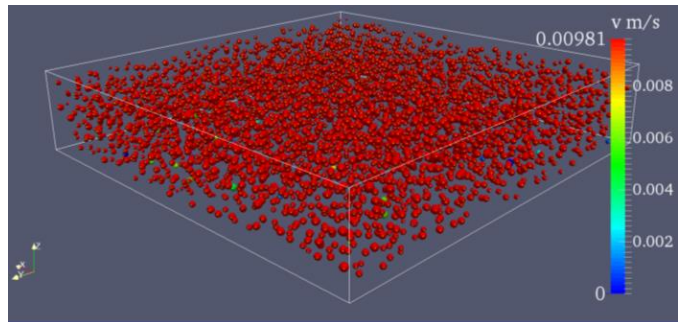
(a) Particles at t = 1×10⁻⁸ sec

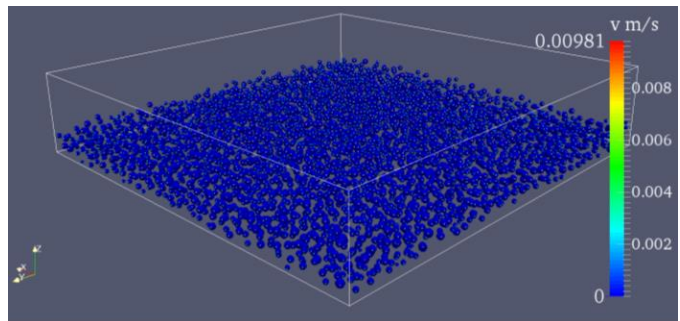
(b) Particles at t = 0.2 sec

Figure 4     Initial and final structure for Gaussian particles from Gran-Hooke-History model with cohesion.

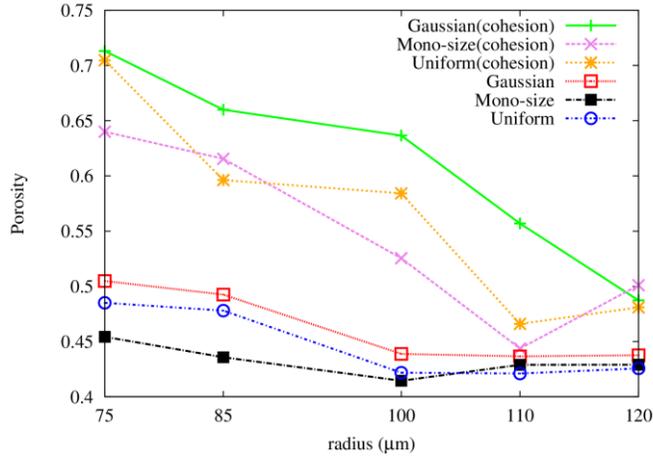
(a) Modified Gran-Hertz-History Model

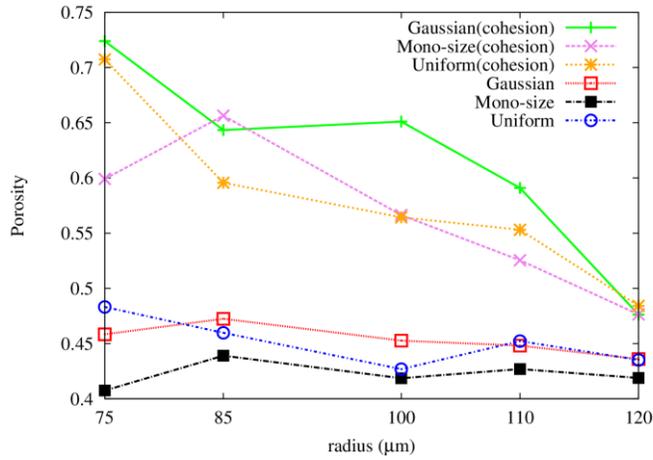
(b) Gran-Hooke-History Model

Figure 5    Effect of porosity with particle size and distribution

Figures 5 and 6 present the porosities and coordination numbers for different cases. It can be seen that the porosity decreases along with the increasing particle radius for all distributions when cohesive forces are considered. Similar trend was observed in the work of previous researchers [30]. This decrease in porosity with increase in radius is expected since with increase of radii or masses of the particles the initial supplied energy (gravitational potential) also increases. So the effect of cohesion in the packing of particles decreases and the porosity values become closer to that for Random Loose Packing [2, 31]. This also explains the decrease in differences between different size distributions in terms of porosity when the radius increases. Among the three distributions considered, Gaussian distribution has the highest porosity and mono-size has the lowest. The porosity values for the two models, Modified Gran-Hertz and Gran-Hooke are slightly different but both show the same trend. As for the non-cohesion case porosity also decreases with increase in particle radius, but the porosity values are much smaller. Figure 5 also shows that the rate of decrease of porosity with radius for non-cohesion case is much smaller. For mono-sized distribution without cohesion, porosity remains almost constant for both Modified Gran-Hertz-History model and Gran-Hooke-History model. Since there is no cohesion the dissipative forces are smaller and particles can pack more closely. Again the difference between the two models in non-cohesion cases is very small. For the coordination number, the trends for three distributions with cohesion are similar. It can be observed that the coordination number increases as particle radius increases which is exactly the opposite of the trend of porosity. Unlike porosity, Gaussian distribution now has the lowest coordination number and mono-size distribution has the highest. Interestingly, it is found that there is no significant change in coordination number whether or not cohesion is included. However one can expect that coordination number should be smaller when there is no cohesion (porosity is larger). This can be explained as follows. When there is cohesion, particles tend to clump together and form clusters. These clusters have void spaces in them. Due to this formation of clusters in some region particles have high coordination number and in some region the coordination number is small. The coordination numbers given in Tables 2-5 and Figure 6 are average of coordination numbers for all particles. It can be seen that the coordination numbers for cohesion and non-cohesion cases are similar.

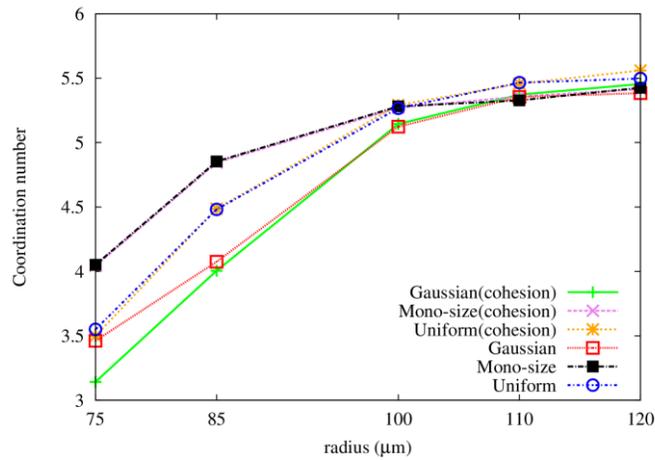

(a) Modified Gran-Hertz-History Model

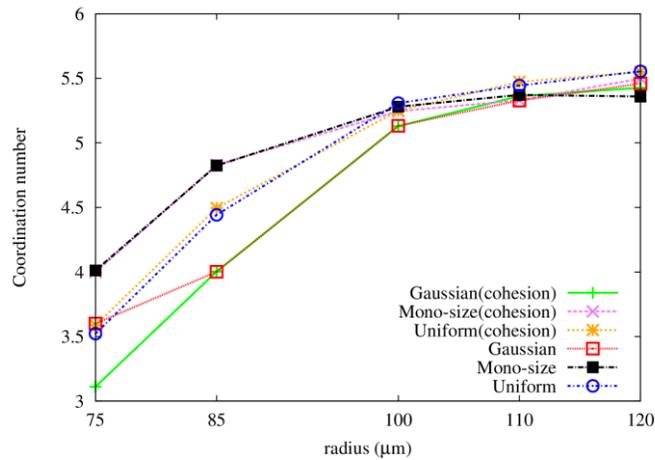

(b) Gran-Hooke-History Model

Figure 6      Effect of coordination number with particle size and distribution

Table 2 Porosity and coordination number for Modified Gran-Hertz-History model

| Radius | Porosity | | | Coordination number | | |
|---|---|---|---|---|---|---|
| | Mono-sized | Uniform | Gaussian | Mono-sized | Uniform | Gaussian |
| 75μm | 0.615 | 0.695 | 0.713 | 4.04 | 3.50 | 3.14 |
| 85μm | 0.656 | 0.685 | 0.660 | 4.84 | 4.49 | 4.01 |
| 100μm | 0.583 | 0.615 | 0.637 | 5.27 | 5.29 | 5.14 |
| 110μm | 0.575 | 0.505 | 0.557 | 5.41 | 5.56 | 5.37 |
| 120μm | 0.485 | 0.574 | 0.487 | 5.36 | 5.46 | 5.45 |

Table 3 Porosity and coordination number for Gran-Hooke-History model

| Radius | Porosity | | | Coordination number | | |
|---|---|---|---|---|---|---|
| | Mono-sized | Uniform | Gaussian | Mono-sized | Uniform | Gaussian |
| 75μm | 0.599 | 0.695 | 0.724 | 4.00 | 3.59 | 3.11 |
| 85μm | 0.656 | 0.685 | 0.643 | 4.83 | 4.50 | 4.00 |
| 100μm | 0.566 | 0.615 | 0.476 | 5.25 | 5.25 | 5.13 |
| 110μm | 0.525 | 0.505 | 0.591 | 5.50 | 5.55 | 5.43 |
| 120μm | 0.476 | 0.574 | 0.651 | 5.33 | 5.47 | 5.36 |

Table 4 Porosity and coordination number for Modified Gran-Hertz-History model without cohesion

| Radius | Porosity | | | Coordination number | | |
|---|---|---|---|---|---|---|
| | Mono-sized | Uniform | Gaussian | Mono-sized | Uniform | Gaussian |
| 75μm | 0.454 | 0.485 | 0.505 | 4.05 | 3.55 | 3.46 |
| 85μm | 0.436 | 0.478 | 0.493 | 4.85 | 4.48 | 4.08 |
| 100μm | 0.415 | 0.422 | 0.439 | 5.28 | 5.27 | 5.12 |
| 110μm | 0.430 | 0.421 | 0.437 | 5.33 | 5.47 | 5.35 |
| 120μm | 0.429 | 0.426 | 0.438 | 5.43 | 5.50 | 5.38 |

Table 5 Porosity and coordination number for Gran-Hooke-History model without cohesion

| Radius | Porosity | | | Coordination number | | |
|---|---|---|---|---|---|---|
| | Mono-sized | Uniform | Gaussian | Mono-sized | Uniform | Gaussian |
| 75μm | 0.407 | 0.483 | 0.458 | 4.01 | 3.52 | 3.60 |
| 85μm | 0.439 | 0.460 | 0.472 | 4.82 | 4.44 | 4.00 |
| 100μm | 0.419 | 0.427 | 0.453 | 5.28 | 5.31 | 5.13 |
| 110μm | 0.427 | 0.452 | 0.448 | 5.37 | 5.44 | 5.33 |
| 120μm | 0.419 | 0.435 | 0.436 | 5.36 | 5.55 | 5.46 |

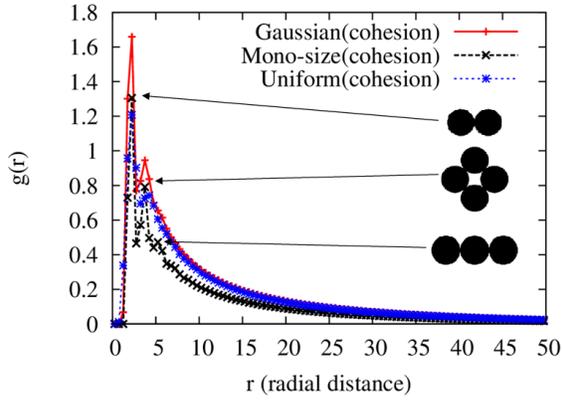

(a) Modified Gran-Hertz-History with cohesion

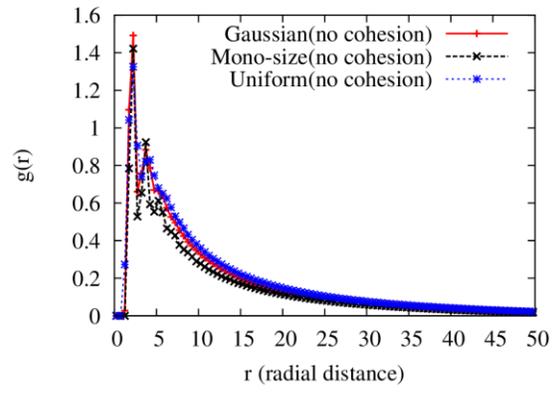

(b) Modified Gran-Hertz-History without cohesion

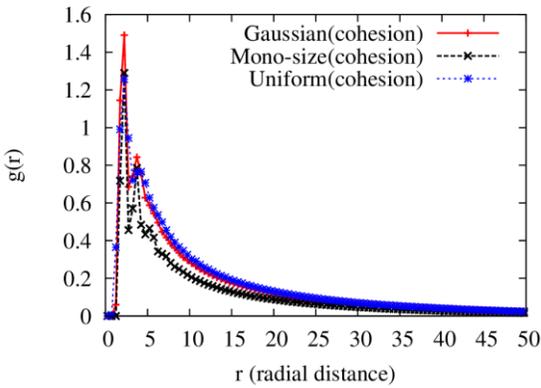

(c) Gran-Hooke-History with cohesion

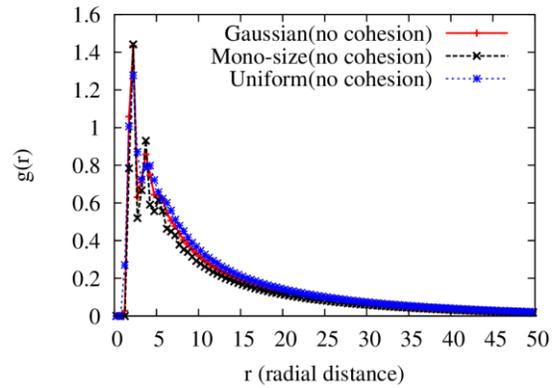

(d) Gran-Hooke-History without cohesion

Figure 7　　RDF for particles with 75 μm radius.

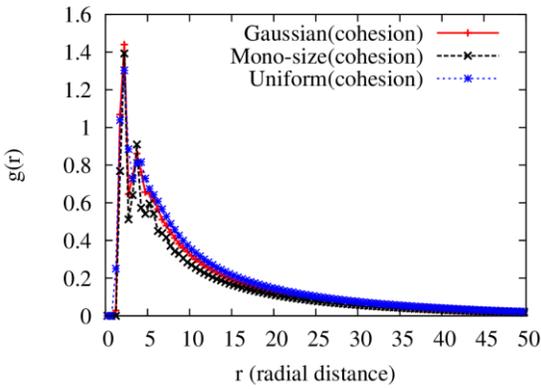

(a) Modified Gran-Hertz-History with cohesion

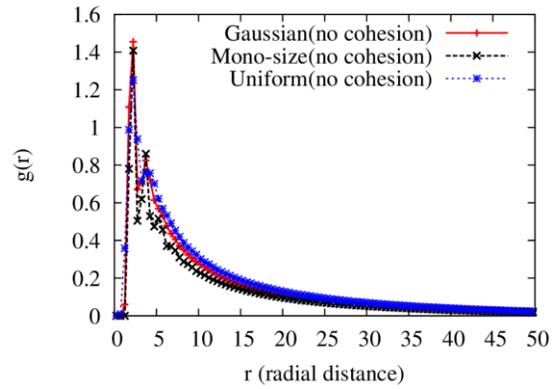

(b) Modified Gran-Hertz-History without cohesion

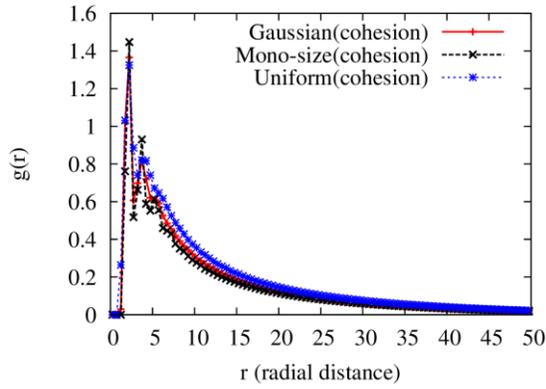
(c) Gran-Hooke-History with cohesion

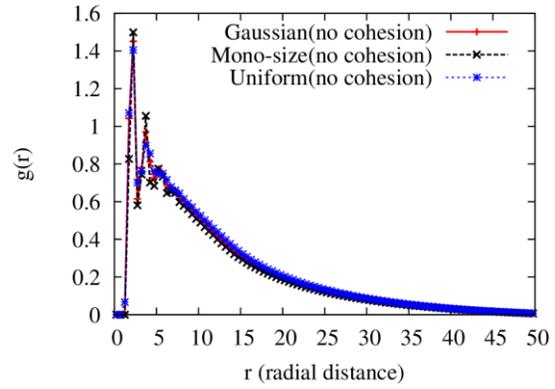
(d) Gran-Hooke-History without cohesion

Figure 8    RDF for particles with 85μm radius.

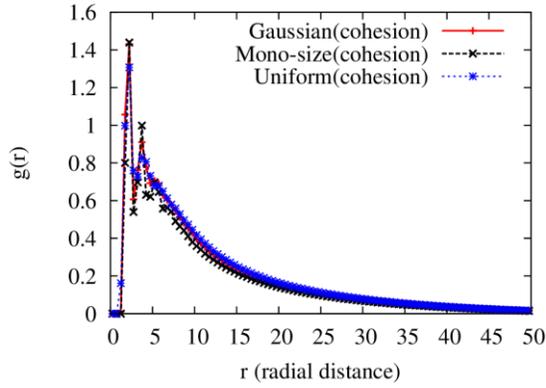
(a) Modified Gran-Hertz-History with cohesion

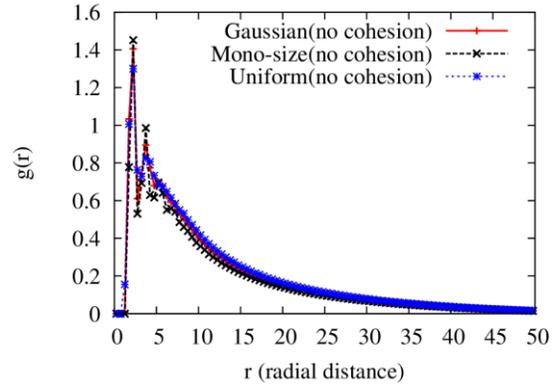
(b) Modified Gran-Hertz-History without cohesion

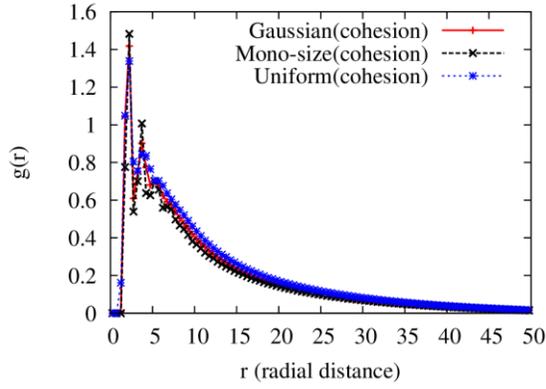
(c) Gran-Hooke-History with cohesion

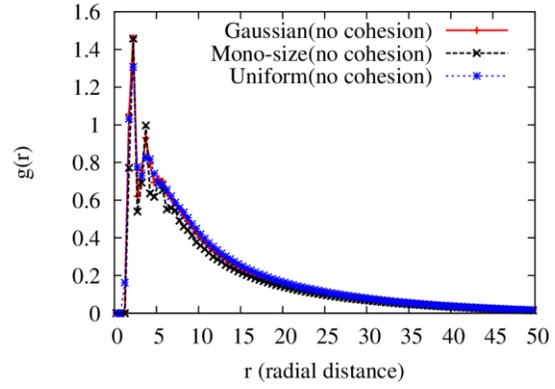
(d) Gran-Hooke-History without cohesion

Figure 9    RDF for particles with 100 μm radius.

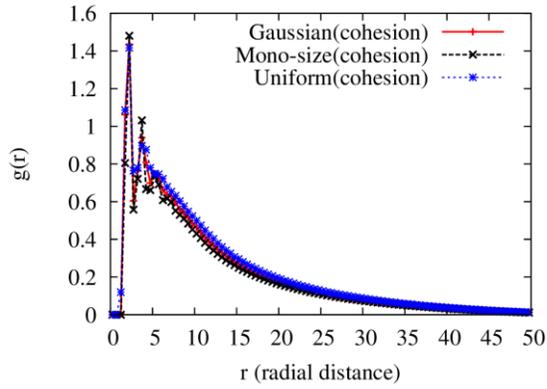
(a) Modified Gran-Hertz-History with cohesion

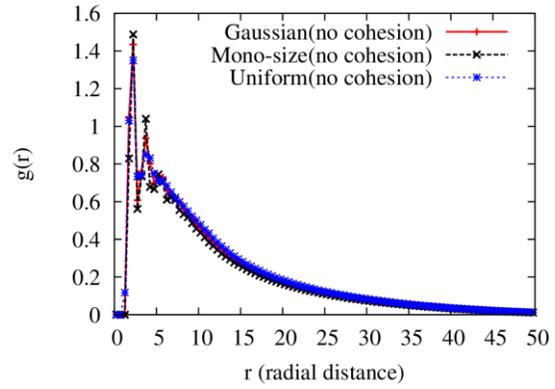
(b) Modified Gran-Hertz-History without cohesion

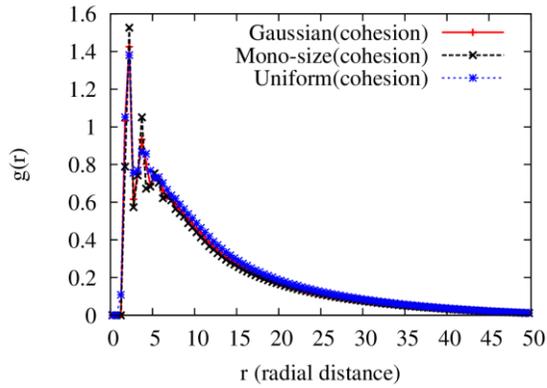
(c) Gran-Hooke-History with cohesion

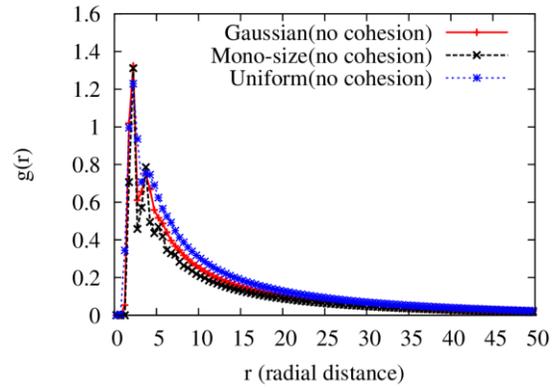
(d) Gran-Hooke-History without cohesion

Figure 10    RDF for particles with 110μm radius.

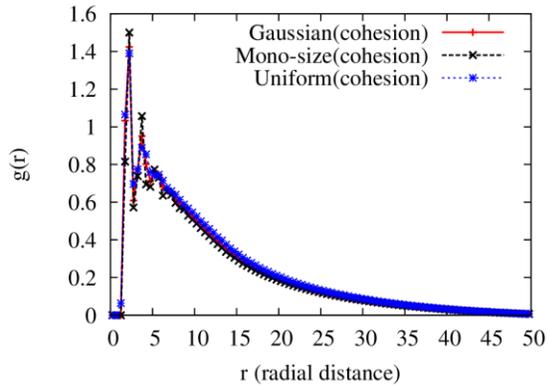
(a) Modified Gran-Hertz-History with cohesion

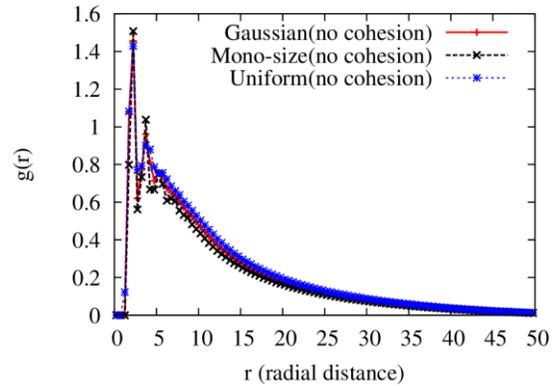
(b) Modified Gran-Hertz-History without cohesion

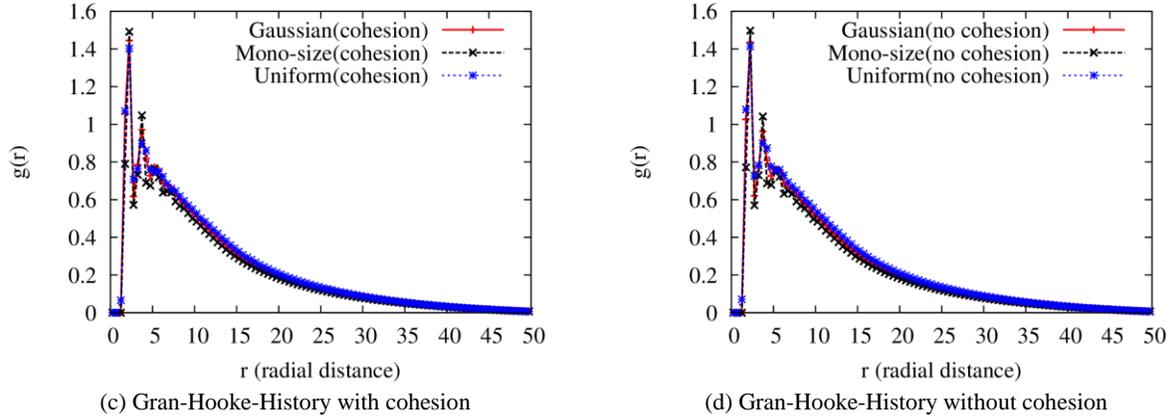

(c) Gran-Hooke-History with cohesion  (d) Gran-Hooke-History without cohesion

Figure 11     RDF for particles with 120μm radius.

Figures 7-11 show the RDF for particle systems with mean radius of 75 μm, 85μm, 100 μm, 110 μm and 120 μm and associated with three different size distributions (mono-sized, uniform and Gaussian). For the cases where the particles have the same radius, three main apparent peaks appear. The first peak is sharply at $2r$ which is for the initial one to one contact, the second and the third are at around $2\sqrt{2}r$ and $4r$, respectively which corresponds to the two characteristic particle contact types, namely edge-sharing-in-plane equilateral triangle and three particles centers in a line (the three contact types are illustrated in Figure 7 (a)). The second and third peaks merge into a single second peak for other distributions. The particle systems with mono-size distribution usually have the highest peak values among all three cases. The peak values for Hertz model are close to that for Hooke model. Also the peak values of RDF are almost same for cohesion and non-cohesion cases.

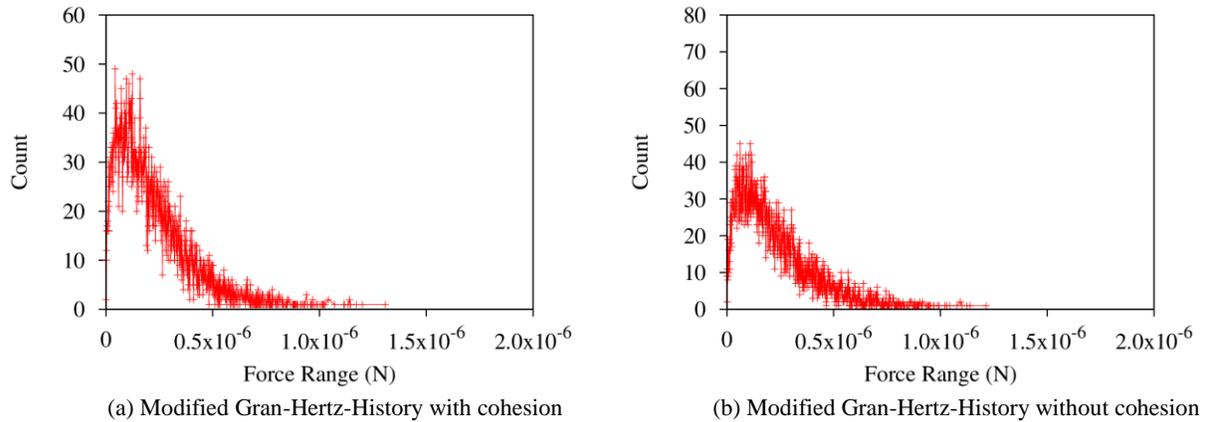

(a) Modified Gran-Hertz-History with cohesion  (b) Modified Gran-Hertz-History without cohesion

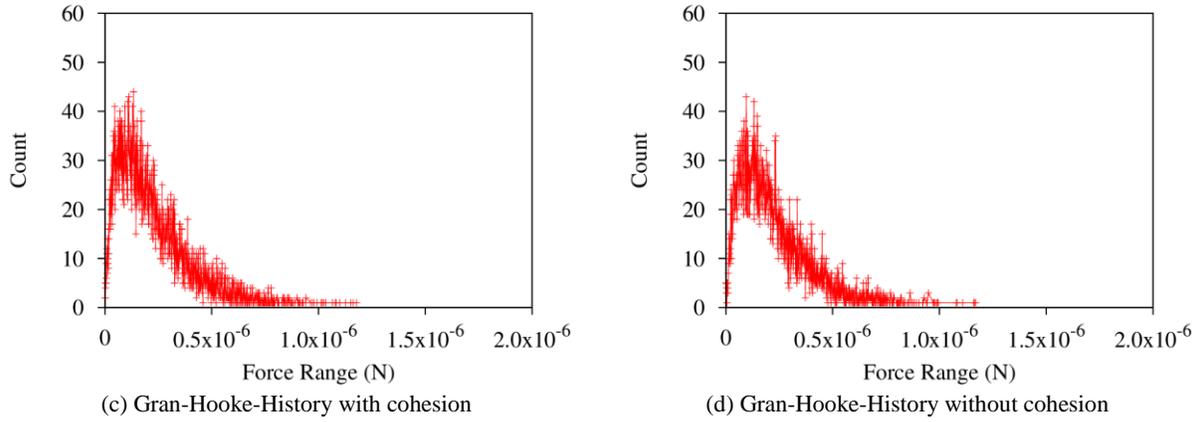

(c) Gran-Hooke-History with cohesion  (d) Gran-Hooke-History without cohesion

Figure 12    Force distribution for particles with 75 μm radius and Gaussian distribution.

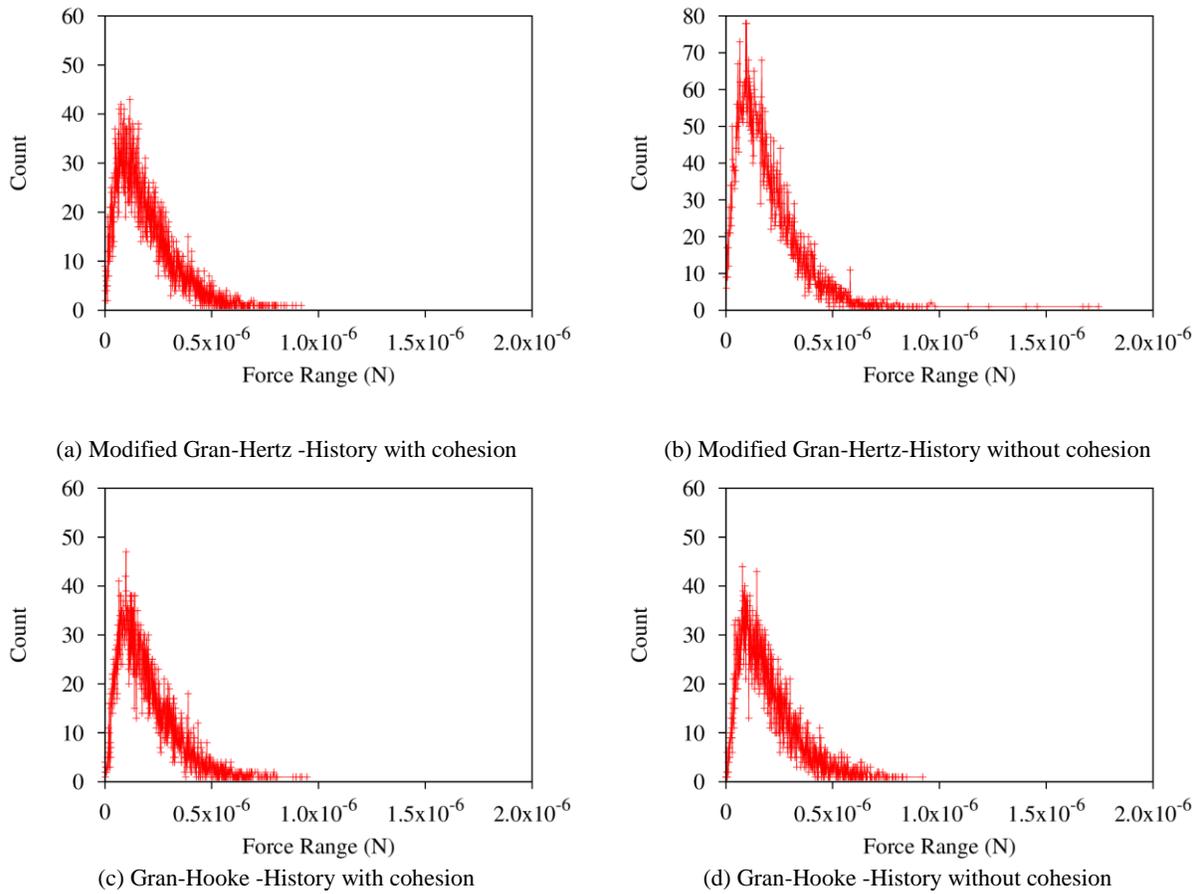

(a) Modified Gran-Hertz -History with cohesion  (b) Modified Gran-Hertz-History without cohesion

(c) Gran-Hooke -History with cohesion  (d) Gran-Hooke -History without cohesion

Figure 13    Force distribution for particles with 75 μm radius and mono-size distribution.

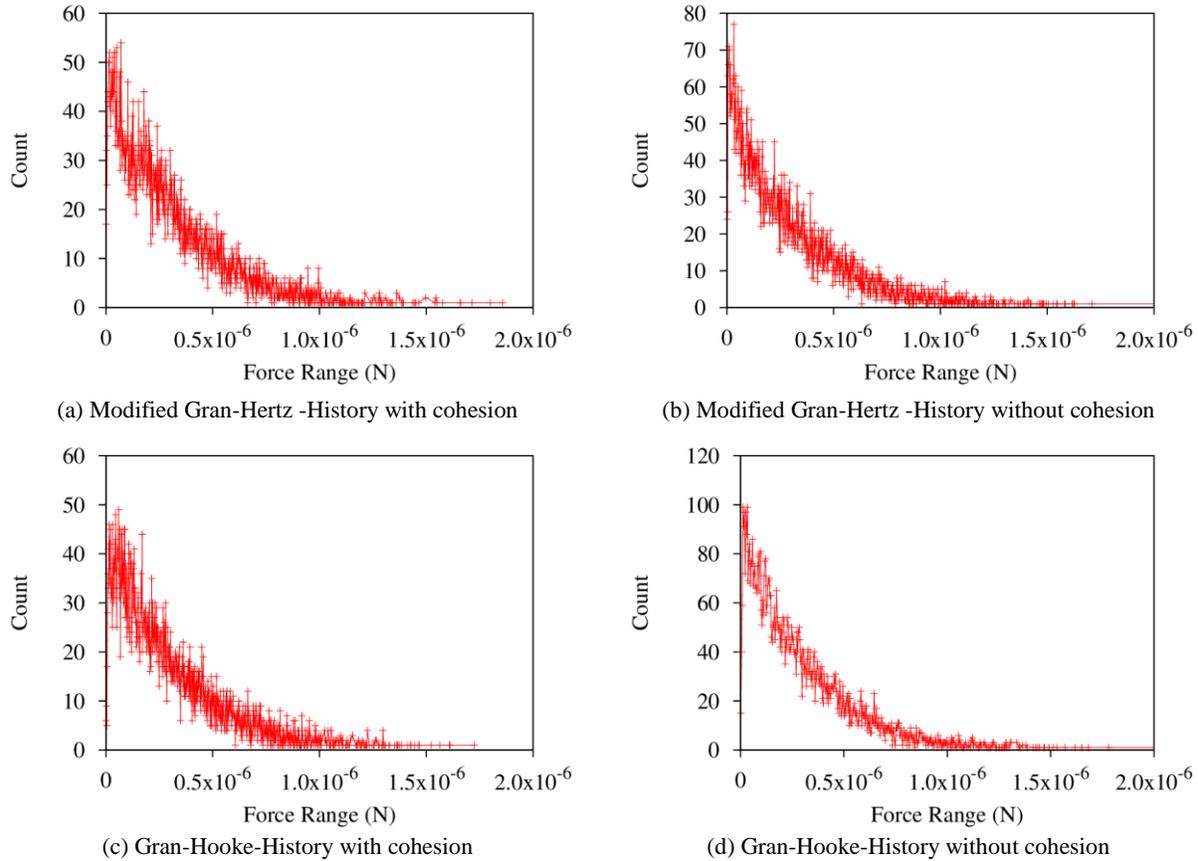

Figure 14    Force distribution for particles with 75 μm radius and uniform distribution.

Figure 12-14 show the force distribution results after the particles are completely packed for particle systems with mean radius of 75μm. The force distribution graphs for other particle radii look similar. For same distribution the counts for each force magnitude are not exactly the same but close. However as the particle radius increases, the force magnitudes increase as a response. Tables 6 and 7 give the mean net force for all the cases when the particles are finally packed. It has to be pointed out that the resultant force here does not represent gravity, since effect of gravity is small (of the order of $10^{-12}$ N). It can also be observed that the net force does not vary much even when the cohesion is included. The mean net force increases with the size of the particles, and it can also be seen that this force has the largest value if the particle size follow uniform distribution. Particles with Gaussian distribution have the secondary magnitude of force, while the mono sized particles have the smallest net force. The difference in magnitude of mean net force between the two different contact models is negligible.

By comparing the two models, Modified Gran-Hertz-History and Gran-Hooke-History, it can be seen that the difference between them is not significant in terms of porosity, coordination number and mean net force. Both of these models assume that the particles are viscoelastic and have a stiffness term and dissipation term. As pointed out by [32] the linear Gran-Hooke model can be as accurate as the non-linear Modified Gran-Hertz model if the stiffness constants, $K_n$ and $K_t$, and damping coefficients, $\gamma_n$ and $\gamma_t$, are evaluated carefully. In this study, even though cohesion is included, the results obtained from the two models are still close. Van der Waals force included in the Modified Gran Hertz model did not seem to play a great role in the packing process. This might be because particle sizes are too large for van der Waals force to take effect. When the efficiency of the two models are considered, the simulations with the Gran-Hooke-History model ran faster than the simulation with the Modified Gran-Hertz-History model. So the linear Gran-Hooke-History model is more efficient than the Modified Gran-Hertz-History model.

Table 6 Magnitude of mean net contact force (N) for Modified Gran-Hertz-History model

| Radius | Cohesion | | | No Cohesion | | |
|---|---|---|---|---|---|---|
| | Mono-sized | Uniform | Gaussian | Mono-sized | Uniform | Gaussian |
| 75μm | $1.91\times10^{-7}$ | $3.00\times10^{-7}$ | $2.20\times10^{-6}$ | $1.94\times10^{-7}$ | $2.87\times10^{-7}$ | $2.25\times10^{-7}$ |
| 85μm | $3.75\times10^{-7}$ | $4.86\times10^{-7}$ | $4.13\times10^{-7}$ | $3.76\times10^{-7}$ | $4.74\times10^{-7}$ | $4.22\times10^{-7}$ |
| 100μm | $8.50\times10^{-7}$ | $1.07\times10^{-6}$ | $9.22\times10^{-7}$ | $8.60\times10^{-7}$ | $1.03\times10^{-6}$ | $9.09\times10^{-7}$ |
| 110μm | $1.41\times10^{-6}$ | $1.59\times10^{-6}$ | $1.52\times10^{-6}$ | $1.41\times10^{-6}$ | $1.59\times10^{-6}$ | $1.50\times10^{-6}$ |
| 120μm | $2.30\times10^{-6}$ | $2.41\times10^{-6}$ | $2.30\times10^{-6}$ | $2.23\times10^{-6}$ | $2.40\times10^{-6}$ | $2.30\times10^{-6}$ |

Table 7 Magnitude of mean net contact force (N) for Gran-Hooke-History model

| Radius | Cohesion | | | No Cohesion | | |
|---|---|---|---|---|---|---|
| | Mono-sized | Uniform | Gaussian | Mono-sized | Uniform | Gaussian |
| 75μm | $1.96\times10^{-7}$ | $2.93\times10^{-7}$ | $2.23\times10^{-7}$ | $2.00\times10^{-7}$ | $2.85\times10^{-7}$ | $2.23\times10^{-7}$ |
| 85μm | $3.92\times10^{-7}$ | $4.94\times10^{-7}$ | $4.30\times10^{-7}$ | $3.94\times10^{-7}$ | $5.46\times10^{-7}$ | $4.10\times10^{-7}$ |
| 100μm | $8.75\times10^{-7}$ | $1.00\times10^{-6}$ | $9.28\times10^{-7}$ | $8.61\times10^{-7}$ | $1.03\times10^{-6}$ | $8.87\times10^{-7}$ |
| 110μm | $1.50\times10^{-6}$ | $1.59\times10^{-6}$ | $1.45\times10^{-6}$ | $1.39\times10^{-6}$ | $1.58\times10^{-6}$ | $1.49\times10^{-6}$ |
| 120μm | $2.24\times10^{-6}$ | $2.35\times10^{-6}$ | $2.31\times10^{-6}$ | $2.19\times10^{-6}$ | $2.44\times10^{-6}$ | $2.48\times10^{-6}$ |

**Conclusions**

A study on packing structures of particle system with different radii and size distributions using two different models are carried out by the Discrete Element Method. The simulation results including RDF and force distribution, porosity and coordination number are presented. It was observed that the particles with Gaussian distribution always have the lowest packing density while the particles with uniform size distribution have the medium packing density and mono-sized particles normally have the highest packing density. For the particles packing under cohesive effect, size distributions result in the same tendency of packing density but has much less variation with particle size. Coordination number is not affected by cohesion significantly but particle size and size distribution do influence the result. The differences in porosity, coordination number, RDF and magnitude of mean net force between the two models used are not substantial which show that any of the models can be used for simulation of particle packing. However when efficiency is considered the Gran-Hooke-History model is found to be more efficient than the Modified Gran-Hertz-History model. So Gran-Hooke-History model can be the model of choice for simulating micro-sized particles.

**Acknowledgement**

Support for this work by the U.S. National Science Foundation under grant number CBET-1404482 is gratefully acknowledged.